\documentclass[aps,twocolumn,prl,showpacs,superscriptaddress]{revtex4-1}

\usepackage{graphicx}% Include figure files
\usepackage{tabularx}
\usepackage{dcolumn}% Align table columns on decimal point
\usepackage{bm}% bold math
\usepackage{amsmath}% More mathematical features
\usepackage{amssymb}
\usepackage{epsfig}
\usepackage{verbatim} % to make comments
\usepackage{pstricks,pst-grad,pst-plot}% PsTricks
\usepackage{natbib}
\definecolor{lightyellow}{cmyk}{0,0,0.3,0}
\definecolor{lightblue}{cmyk}{0.1,0,0,0}
\definecolor{green4}{cmyk}{0.0,0.0,0.7,0.4}
\definecolor{dgreen}{rgb}{0,.4,0}
\definecolor{yellow2}{rgb}{.7,.7,0}

\newcommand{\invisible}[1]{}
\begin{document}

%\title{Epidemics in Soil Networks}
\title{Prominent effect of soil network heterogeneity on microbial
invasion}

%\begin{comment}
\author{F.J.~P{\'e}rez-Reche}
\affiliation{SIMBIOS Centre, University of Abertay,  Dundee,  UK}
%\email{p.perezreche@abertay.ac.uk}

\author{S.N.~Taraskin}
\affiliation{St. Catharine's College and Department of Chemistry,
University of Cambridge, Cambridge, UK}

\author{W. Otten}
\affiliation{SIMBIOS Centre, University of Abertay,  Dundee,  UK}

\author{M.P. Viana}
\affiliation{Instituto de Fisica de Sao Carlos, Universidade de Sao Paulo, Sao
Carlos, SP, Brazil}

\author{L. da F. Costa}
\affiliation{Instituto de Fisica de Sao Carlos, Universidade de Sao Paulo, Sao
Carlos, SP, Brazil}

\author{C.A.~Gilligan}
\affiliation{Department of Plant Sciences, University of Cambridge,
Cambridge,  UK}
%\end{comment}

\begin{abstract}

Using a network representation for real soil samples and
mathematical models for microbial spread, we show that the structural
heterogeneity of the soil habitat may have a very significant influence on the size of
microbial invasions of the soil pore space. In particular, neglecting the soil structural heterogeneity
may lead to a substantial underestimation of microbial invasion. Such
effects are explained in terms of a crucial interplay between
heterogeneity in microbial spread and heterogeneity in the topology of
soil networks. The main influence of network topology on invasion is linked to the existence of long channels in soil networks that may act as bridges for transmission of microorganisms between distant parts of soil.

\end{abstract}

%\pacs{}
%87.23.Cc 	Population dynamics and ecological pattern formation 
%89.75.Hc 	Networks and genealogical trees 
%05.10.-a 	Computational methods in statistical physics and nonlinear dynamics
%89.60.-k 	Environmental studies
\maketitle

\date{\today}

%%%Introduction
Understanding how the ubiquitous structural heterogeneity of
  natural habitats affects the movement and spatial distribution of
  biota is an important and fascinating question relevant to several 
disciplines~\cite{Turner_Book2001_LandscapeEcology,Viswanathan_Stanley_Book:PhysicsForaging}. 
In  particular, the soil pore space is a highly heterogeneous habitat
  hosting a stunning wealth of biological activity (e.g. of bacteria,
  fungi or
  nematodes~\cite{Lavelle_Book2003,Copley_Nature2000,Young_Crawford_Science2004})
  that plays an essential role in many processes including
 plant growth~\cite{Kiers_Science2011}, climate
change~\cite{Singh_NatureRev2010}, or soil-borne
epidemics~\cite{Otten_EurJSoilScience2006}.  
The study of the
interplay between soil structural heterogeneity and microbial
activities in three dimensions (3D) is challenging due to the opacity
of soil and the complexity of biological and environmental factors
involved in microbial spread.  
Experiments based on soil thin
sections~\cite{Franklin_2007,Otten_NewPhytol1998,Nunan_MicrobialEcology2002,Harris_FEMSMicrobiolEcol2003,Otten_SoilBiolBiochem2004}
  or planar
  microcosms~\cite{Hapca_JRoySocInterface2009,Dechesne_PNAS2010,Franklin_2007}
give some insight.
For instance, it was observed that the volume
  of soil explored by fungi increases with the soil bulk
  density~\cite{Harris_FEMSMicrobiolEcol2003} and macropores may act
  as either preferential pathways or barriers for fungal
  spread~\cite{Otten_SoilBiolBiochem2004}. 
However, due to the nature of
  the techniques, these type of experiments fail to provide 
information in 3D so as to quantitatively assess the
  influence of the structural heterogeneity and topology on
  microbial invasion.
Current understanding based on ecological and
epidemiological models suggests that heterogeneity in soil structure could
either enhance or reduce the probabilities of
invasions~\cite{Melbourne_EcologyLetters2007,Neri_JRSInterface2010,Newman_PRE2002,Miller_PRE2007}. 
The outcome
depends largely on the properties of the pore space, including the connectivity
and pore sizes, and the effects these properties have on
microbial movement through soil. 
In this letter, we identify the main structural 
factors that affect invasion by devising several network models for
biological invasion
with increasing degree of interplay between microbial spread and the
structure of the soil pore space (Table~\ref{Table_0}). 
Our results and conclusions are not only relevant to biological invasion in soil but are also expected to be important for any biota moving in complex landscapes or generic agents spreading in networks with structurally complex links.

%%%%Methods
We have analysed the invasion models for six real soil samples: three samples of
soil without
tillage treatment (denoted as N1, N2, and N3) and three
samples of ploughed soil (P1, P2, and P3).
The main qualitative difference between N and P
samples is that the pores are typically larger in the P
samples (see Fig.~\ref{fig:Network}(a) and the statistical analysis in \cite{Supplemental_HeteroSoil}).
For our theoretical analysis, we have used a network representation
of the soil pore
structure derived from 3D-digital images of the soil samples scanned with an
X-ray
micro-tomography device \cite{Supplemental_HeteroSoil}.
Soil networks consist of a set of nodes and edges whose layout captures the
topology
of the soil pore space where biological activity takes place.
The network representation is achieved by associating the branching points of the soil pore space
with the nodes and the pore-space channels between branching points with the 
network edges (see \cite{Supplemental_HeteroSoil} for more detail and a comparison with previously proposed network representations for soil~\cite{Vogel_EJSS1998,Vogel_AdvWatRes2001,Santiago_NonlinProcessesGeophys2008,Cardenas_Geoderma2010}).
Fig.~\ref{fig:Network}(b) shows the network for sample
N1.
All the networks have similar topological properties irrespective
of tillage treatment
(Table~\ref{Table_1}). The samples exhibit limited node degree, small topological
heterogeneity, high clustering in comparison with random graphs, and
\emph{fractal} small-world behaviour
(i.e. the mean separation between nodes increases as
$\langle l \rangle \sim N^{\eta}$ with $\eta \simeq 0.4$, which contrasts 
with the slower increase $\langle l \rangle \sim \ln N$ in standard small-wold
networks ~\cite{Csanyi_PRE2004}). 
Such features are typical for geographical
networks embedded in Euclidean spaces~\cite{Boccaletti:2006,Buhl_Sole_EPJB2004}
and are
thus not un-expected for soil networks that are embedded in a 3D space.
The connectivity in this kind of networks is limited because each edge fills a
certain space and thus the number of edges per node is
restricted~\cite{Boccaletti:2006,Buhl_Sole_EPJB2004,Supplemental_HeteroSoil}.
This property remains unaltered under tillage and
this is likely to be the reason why the topologies are statistically
similar for both N and P samples.
We describe the structure of channels in terms of their arc-length,
$L$, and local cross-section area $S(x)$ (Fig.~\ref{fig:Network}).
Both the mean value  $\langle L \rangle$ and the relative dispersion 
$\sigma_\text{L}^2/\langle L \rangle^2$ of the arc-length are similar for all the samples
(Table~\ref{Table_1}). 
A relatively weak  heterogeneity in $L$, 
i.e. $\sigma_\text{L}^2/\langle L \rangle^2 < 1$, contrasts with much greater variability in
 the cross-section area along channels, i.e. $\sigma_\text{S}^2/\langle S \rangle^2 > 1$. 
Large cross-sections are typical for long channels
which are frequently attached to nodes with large $k$. Spatial correlations of $S(x)$ are significant (i.e. narrow channels tend to be attached to narrow channels and vice-versa)~\cite{Supplemental_HeteroSoil}.
Here, we show that the heterogeneity in $S(x)$, the correlations between $L$ and $S(x)$ and spatial correlations for $S(x)$ play a key role for microbial invasion.

\begin{table}[p]
\caption{\label{Table_0} Models for microbial spread. 
$T$ is the transmissibility. $L$ ($S$) is the arc-length (cross-section) of channels. $\lambda_0$ ($\alpha$) is the length (area) exploration parameter. $\overline{S^{\alpha}}=L^{-1} \int_0^L [S(x)]^\alpha \text{d}x$ is the average of $S(x)^{\alpha}$ along channels. Derivations of $T$ for models 3 and 4 are
presented in \cite{Supplemental_HeteroSoil}.}
%\vspace{10pt}
%\begin{tabular}[t]{c!{ }c!{ }c!{ }c!{ }c!{ }c!{ }c!{ }c!{ }c!{ }c!{ }c}
\begin{tabular}{c!{ }c!{ }c!{ }c}
\hline
Model & Heterogeneity & Parameters & Transmissibility\\
\hline
1 & None & $T$ & $T$\\
2 & Non-structural & $\langle T \rangle$ &
$
\begin{cases}
0,&\text{with Prob } 1-\langle T \rangle\\
1,&\text{with Prob } \langle T \rangle
\end{cases}
$\\
3 & $L$ & $\lambda_0$ & $\exp(-L/\lambda_0)$\\
4 & $L,\; S(x)$ & $\lambda_0,\; \alpha$ & $\exp(-L\,
\overline{S^{\alpha}}/\lambda_0)$\\
\hline
%\end{tabular}
\end{tabular}
\end{table}

%%%%%%%%%%%%%%%%%%%%%%%%%%%%%%%%%%%%%%%%%%%%%%%%%%%%%
%%%%%%%%%%%%%%%%% Figure 1 %%%%%%%%%%%%%%%%%%%%%%%%%%
%%%%%%%%%%%%%%%%%%%%%%%%%%%%%%%%%%%%%%%%%%%%%%%%%%%%%
\begin{figure}%[p]
\begin{center}

{\includegraphics[clip=true,width=8cm]{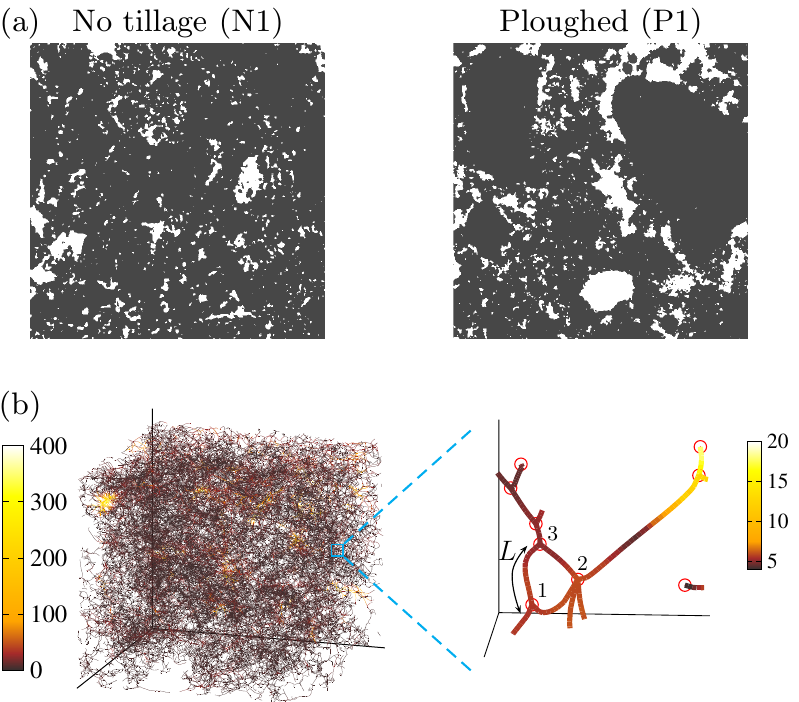}}

\end{center}
\caption{ \label{fig:Network}Soil samples and soil networks. (a) 2D-sections of
3D-digital images of soil samples N1 and P1 (sample size: $3.5 \times 3.5 \times
3.5$~cm). The pore space and solid matrix are shown in white and
grey colours, respectively. 
(b) Network representing the pore space of sample
N1. The colour of channels indicates their local cross-section area, $S(x)$,
where $x$ gives the position along channels.
The zoom on the right illustrates the definition of the
arc-length of channels, $L$, the node degree heterogeneity, and
clustering (triangle 1-2-3) (cf. Table~\ref{Table_1}). 
Both $L$ and $S(x)$
are dimensionless, i.e. scaled by the size of voxel in the 3D-digital image. 
% of soil.
}
\end{figure}

\begin{table*}%[p]
\caption{\label{Table_1} Network topological characteristics and channel
properties for six soil samples. $N$ and $E$ denote the number of nodes and the
number of edges in networks, respectively. $\langle k \rangle$ is the mean
degree and $\sigma_k^2/\langle k \rangle^2= \langle k^2 \rangle/\langle k
\rangle^2 - 1$ is a measure of the
topological heterogeneity~\cite{Barrat_08:book}.
 The clustering $C$ is given relative to $C_{\text{rand}}
= \langle k \rangle/N$ for a random graph with the same value of $N$
and $\langle k \rangle$~\cite{Watts_Strogatz:1998}.
 The mean separation length, $\langle l \rangle$, 
gives the typical separation
between two nodes in the network~\cite{Watts_Strogatz:1998}. $\langle L
\rangle$  and $\sigma_\text{L}^2/\langle L \rangle^2$ are the mean value and
dispersion of $L$. Analogous quantities for cross-section area $S$ are given in
the last two columns.}
%\vspace{10pt}
%\begin{tabular}[t]{c!{ }c!{ }c!{ }c!{ }c!{ }c!{ }c!{ }c!{ }c!{ }c!{ }c}
\begin{tabular}{c!{ }|c!{ }c!{ }c!{ }c!{ }c!{ }c!{ }|!{ }c!{ }c!{ }c!{ }c}
\hline
Sample & \multicolumn{6}{c}{Topological characteristics} &
\multicolumn{4}{c}{Channel properties}\\
 & $N$ & $E$ & $\langle k \rangle$ & $\sigma_k^2/\langle k \rangle^2$&
$C/C_{\text{rand}}$ & $\langle l \rangle $ & $\langle L \rangle$ &
$\sigma_\text{L}^2/\langle L \rangle^2$ & $\langle S \rangle$ &
$\sigma_\text{S}^2/\langle S \rangle^2$\\
\hline
N1 & 49709 & 69563 & 2.80 & 0.165 & 763.7 & 73.05 & 6.47 & 0.362 & 12.03 & 2.51
\\
N2 & 58618 & 82949 & 2.83 & 0.180 & 1000.4 & 74.67 & 5.94 & 0.348 & 11.53 & 6.66
\\
N3 & 54083 & 76747 & 2.84 & 0.165 & 848.0 & 64.26 & 6.53 & 0.382 & 22.80 & 30.60
\\
P1 & 33526 & 45544 & 2.72 & 0.162 & 488.2 & 69.61 & 6.85 & 0.401 & 19.18 & 3.23
\\
P2 & 47388 & 66147 & 2.79 & 0.156 & 667.6 & 66.65 & 6.67 & 0.366 & 17.33 & 5.27
\\
P3 & 27042 & 36125 & 2.67 & 0.165 & 368.3 & 70.73 & 7.15 & 0.450 & 21.31 & 4.00
\\
\hline
%\end{tabular}
\end{tabular}
\end{table*}

%%%%Model for invasion
The spread of micro-organisms through a given pore space in soil is not a deterministic process 
but
it occurs with certain
probability~\cite{Otten_EurJSoilScience2006,Franklin_2007}. Inspired by
epidemiological
network models~\cite{grassberger1983,Newman_PRE2002,Miller_PRE2007,Neri_JRSInterface2010,Perez_Reche_Synergy2011,Vespignani_NaturePhys2012}, we assume that microorganisms reaching a node in
the soil network are able to colonise any of the channels that emerge from that
node and to reach uncolonised nodes with probability $T$ (referred to as the
transmissibility).
This quantity is central for all the models proposed in this work.
Each model assumes a different form for $T$ (Table~\ref{Table_0}).
\emph{Model 1} corresponds to the simplest mean-field case with $T$ being identical for all the channels.
In \emph{model 2}, $T$ is independent of the
structural properties of channels but it takes a random value for each channel
(representing, e.g. a non-uniform spatial distribution of nutrient resources
necessary for microbial activity~\cite{Ettema_TREE2002_SpatialSoil})
that obeys a bi-modal distribution parameterised by the mean transmissibility,
$\langle T \rangle$.
\emph{Model 3} suggests that $T$ depends on the arc-length of channels ($L$) and the spatial scale of microbial colonisation is characterised by a typical
exploration length, $\lambda_0$. 
The value of transmissibility is assumed to decay with increasing $L$, 
meaning that microbial transmission
through short channels is more likely than through longer channels.
In \emph{Model 4}, $T$ depends on both $L$ and the cross-section
area along channels, $S(x)$. 
The dependence on $L$ is again controlled by the
parameter $\lambda_0$.
Regarding $S(x)$, we keep our description general so as to
account both for microorganisms with preferential spread through pores with wide
cross-sections  and those
that have a preference for narrow cross-sections.
This preference may depend on a combination of biological and physical factors
such as competitive exclusion from pore-size classes due to predator-prey
interactions~\cite{Turner_Book2001_LandscapeEcology,Melbourne_EcologyLetters2007,Lavelle_Book2003}, or the spatial distribution of water in
soil~\cite{Otten_NewPhytol1998,Lavelle_Book2003,Franklin_2007,Dechesne_PNAS2010}. 
We capture
these factors qualitatively with an effective area exploration parameter, $\alpha$,
whose sign
controls the preference for narrow ($\alpha>0$) or wide ($\alpha<0$)
cross-sections.
For $\alpha=0$, model 4 reduces to model 3 (cf. expressions for $T$ in
Table~\ref{Table_0}).

%%%% Results

%%%%%%%%%%%%%%%%%%%%%%%%%%%%%%%%%%%%%%%%%%%%%%%%%%%%%
%% Invasion curves                                 %%
%%%%%%%%%%%%%%%%%%%%%%%%%%%%%%%%%%%%%%%%%%%%%%%%%%%%%
\begin{figure}%[p]
%\begin{center}

\includegraphics[clip=true,width=8cm]{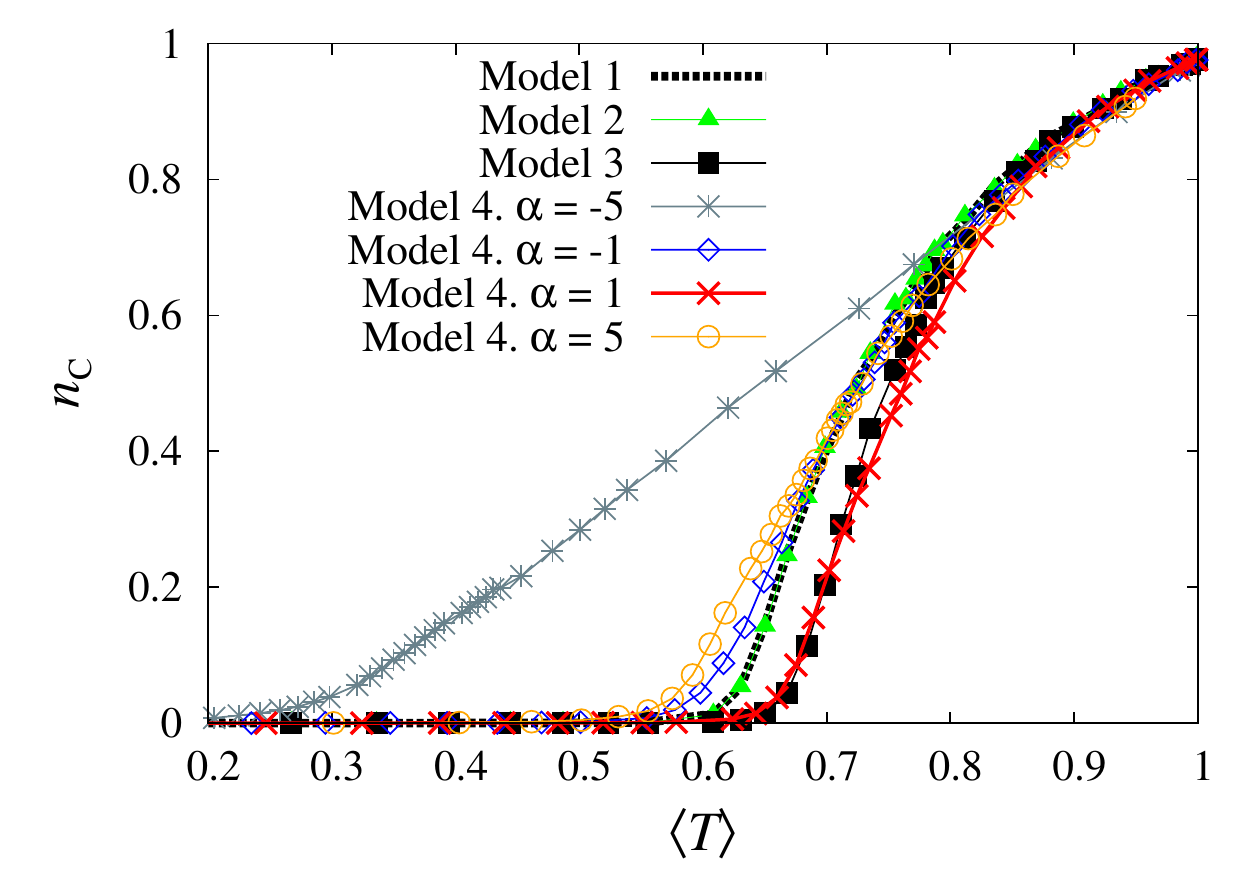}

%\end{center}
\caption{ \label{fig:n_C}
Invasion in soil networks. The vertical axis displays the density
of colonisation, $n_{\text{C}}$, in network N1 \emph{averaged} over $10^4$
invasions starting from
different randomly chosen nodes.
 The horizontal axis shows the mean
transmissibility of microbes, $\langle T \rangle$, obtained by
averaging over channels. Different curves correspond to invasions 
predicted by different
models (and exploration controlled by $\alpha$ in model 4).
}
\end{figure}

We quantify the size of invasion by the density of colonisation,
$n_C$, defined as the relative number of nodes reached by the
microbial colony during an invasion that starts from a randomly chosen
node in the network. 
Numerical simulations reveal that the mean value
of $n_C$ predicted by model 1 increases with $T$ but only takes
significantly large values (e.g. $n_C \gtrsim 0.1$)
if $T>T_{\text{c}}$, where $T_{\text{c}}
\simeq 0.6$ for all the analysed networks (Fig.~\ref{fig:n_C} and \cite{Supplemental_HeteroSoil}).
Similar threshold behaviour for invasion
is typically observed in some epidemiological
models~\cite{grassberger1983,Newman_PRE2002,Miller_PRE2007,Neri_JRSInterface2010,Perez_Reche_Synergy2011,Vespignani_NaturePhys2012}. 
The heterogeneity in $T$
considered in model 2 does not introduce significant differences to
$n_{\text{C}}$ which coincides with that for model 1 if the
strength of microbial transmission is parametrised by $\langle T
\rangle$ in both models.
This result holds in general if the values of $T$ for all the channels in the network are statistically independent from each
other~\cite{sander2002,PerezReche_JRSInterface2010}.
For given $\langle T \rangle$, model 3 predicts colonisations of
smaller size on average than that for model 1 
(the colonisation curve for model
3 is clearly below the curve for model 1 in Fig.~\ref{fig:n_C}).
Therefore, heterogeneity in $T$ induced by heterogeneity in $L$ makes the soil
network more resilient to microbial invasion~\cite{Perez-Reche_IEEE2009}.
Model 4 predicts a similar behaviour for values of $\alpha$ near zero
which is expected since models 4 and 3 coincide for $\alpha=0$ (see,
e.g. the invasion curve for $\alpha=1$ in Fig.~\ref{fig:n_C}). In
contrast, for larger values of the
area exploration parameter (either $\alpha >0$ or $\alpha <0$), invasions for
given $\langle T \rangle$ can be more significant in model 4 than in any of the other models.
The effect is especially pronounced
for large negative $\alpha$ (compare, e.g. the curve for $\alpha=-5$ with that
for homogeneous $T$ in Fig.~\ref{fig:n_C}).
This is a clear illustration of the prominent effect of
the strong heterogeneity in cross-sections of soil channels on biological
invasion.

%%%% Discussion

%%%%%%%%%%%%%%%%%%%%%%%%%%%%%%%%%%%%%%%%%%%%%%%%%%%%%%%%%%%%%%%%%%%%%%
%     Bridgeness given L
%%%%%%%%%%%%%%%%%%%%%%%%%%%%%%%%%%%%%%%%%%%%%%%%%%%%%%%%%%%%%%%%%%%%%%
\begin{figure}%[p]
\begin{center}

{\includegraphics[clip=true,width=8cm]{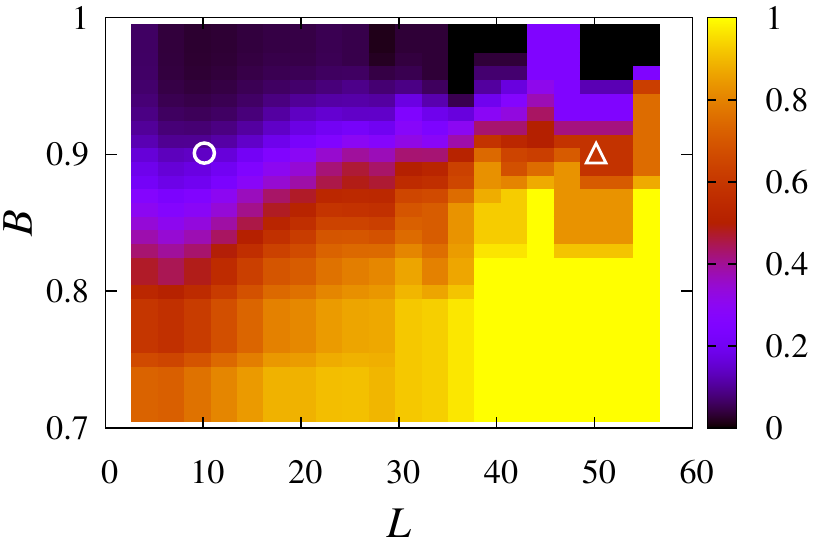}}

\end{center}
\caption{ \label{fig:Bridgeness_L}
Bridging effect for edges in network N1.
The colour along the vertical direction gives the fraction of edges with
given $L$ (i.e. fixed value on the horizontal axis) whose degree of bridging is
larger than the value $B$ plotted along the vertical axis.
For instance, the colour inside the circle (triangle) indicates that the
fraction of
channels of length $L=10$ ($L=50$) with $B>0.9$, i.e.
$l_{\text{b}}>10$, is approximately $0.1$ ($0.7$). By
  definition, $B \in [0,1]$ but only values of $B>0.7$ are shown to
  highlight the variation in $B$ with $L$. 
Channels attached to border nodes of degree 1 are not included in the graph. See
\cite{Supplemental_HeteroSoil} for similar plots for other
soil samples.
}
\end{figure}
%%%%%%%%%%%%%%%%%%%%%%%%%%%%%%%%%%%%%%

%%%%%%%%%%%%%%%%%%%%%%%%%%%%%%%%%%%%%%%%%%%%%%%%%%%%%%%%%%%%%%%%%%%%%%
%     Networks for bonds with <T> > 0.5                              %
%                 P(T|L) for <T>=0.5                                 %
%                                                                    %
%%%%%%%%%%%%%%%%%%%%%%%%%%%%%%%%%%%%%%%%%%%%%%%%%%%%%%%%%%%%%%%%%%%%%%
\begin{figure}%[p]
\begin{center}

{\includegraphics[width=7.5cm]{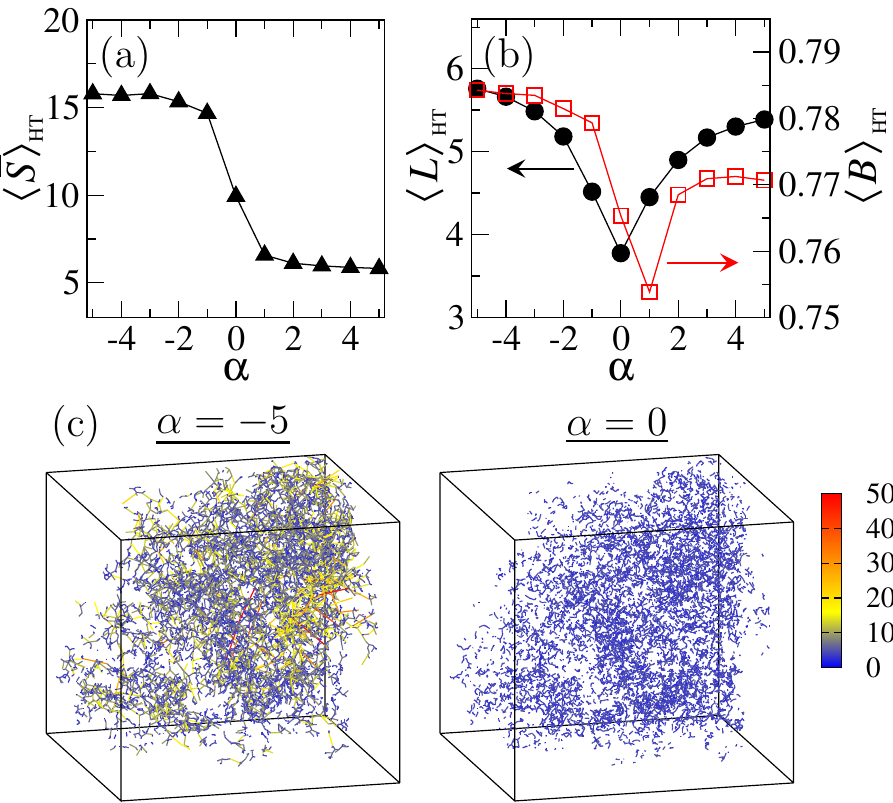}}

\end{center}
\caption{ \label{fig:F_TgivenL}
(a) Mean cross-section area of channels, $\langle \overline{S} \rangle_\text{HT}$, (b) mean arc-length, $\langle L \rangle_\text{HT}$ and mean bridgeness, $\langle B \rangle_\text{HT}$, obtained by averaging over channels with high transmissibility (HT, $T> \langle T \rangle$) in sample N1 for microbes with $\langle T \rangle=0.5$. (c) Plots of the channels with $T>0.5$ coloured according to their value of $L$.
}
\end{figure}
%%%%%%%%%%%%%%%%%%%%%%%%%%%%%%%%%%%%%%

The above results can be understood in terms of the intuitive idea that some
channels act as bridges that link nodes that would
otherwise be further apart or disconnected. Accordingly,
if some microorganisms are able to colonise channels with high
``bridging-effects'', the resulting invasions should be large. 
This behaviour is reminiscent of the small-world effect~\cite{Watts_Strogatz:1998,sander2002} 
and is related to the fractal small-world property of soil networks~\cite{Csanyi_PRE2004}.
In order to quantify the bridge effect for every pair, $u-v$, of connected nodes we measure 
the range of the edge $u-v$ \cite{Motter_PRE2002} which gives
the number of links, $l_{\text{b}}$, along the
shortest path from $u$ to $v$, if the edge linking these two nodes is
removed and define the bridgeness, $B$, as 
$B=1-1/l_{\text{b}}$. 
Channels with  $B$ close to unity indicate the presence of
bridges for
transmission because the shortest alternative path has a large number
of steps, $l_{\text{b}}$.
Fig.~\ref{fig:Bridgeness_L} shows that $B$ is
heterogeneous but is correlated with the channel length,
typically taking
larger values for channels with long
arc-length.  
%For example, $70\%$ of channels of
%length $L=50$ have $B>0.9$ whereas only $10\%$ of channels of length
%$L=10$ have $B>0.9$.
This effect is ultimately responsible for the variations between invasion curves in Fig.~\ref{fig:n_C} for different models.

In models 1 and 2 for invasion, $T$ is not linked to {the} structural properties
of channels and is thus 
independent of $B$. In contrast, model 3 assumes that $T$ decays monotonically
with $L$ meaning that
microorganisms are more likely to be transmitted through short channels which
typically have small $B$ (rather than through channels of any length as in
models 1 and 2).
This explains why the typical size of invasions
for given $\langle T \rangle$ is smaller in model 3 than predicted by
models 1 and 2.
In contrast, the transmissibility in model 4 depends
on $S(x)$ (provided $\alpha \neq 0$) and does not decay monotonically
with $L$ as in model 3 (see Sec. VI in \cite{Supplemental_HeteroSoil}). 
As a result, some long channels with high degrees of bridging are able
to transmit microbes more efficiently than other with smaller $L$ and $B$.
Fig.~\ref{fig:F_TgivenL} illustrates this idea for
microbial invasion
with given
$\langle T \rangle=0.5$ by showing that the cross-section area of highly transmissible channels with $T> \langle T \rangle$ decreases with $\alpha$ (Fig.~\ref{fig:F_TgivenL}(a)) but their length and bridgeness show a minimum for $\alpha \simeq 0$ (Fig.~\ref{fig:F_TgivenL}(b) and (c)). The larger values of $B$ observed for $\alpha<0$ (i.e. when wide channels are preferred) are due to the positive correlations between $S(x)$ and $L$. In spite of these correlations, invasions with positive $\alpha$ (preference for narrow channels) may have a large value of $B$ because $S(x)$ is very heterogeneous and there are channels that are both narrow and long \cite{Supplemental_HeteroSoil}. 
The invasions corresponding to $\alpha \neq 0$ are then
typically larger than predicted by models 1-3 (Fig.~\ref{fig:n_C}). 
The effect is particularly important for $\alpha<0$ because wide channels (preferred for $\alpha<0$) tend to be longer than narrow channels. Similarly, the significant spatial correlations in $S$ favour invasion in model 4 (see \cite{Supplemental_HeteroSoil}).

According to our analysis, the global traits of invasion are mainly dictated
by generic characteristics such as the heterogeneity in the network
(i.e. in channel length and degree of bridging)
and transmission.
An important conclusion which applies to invasions in any
  heterogeneous landscape is that a limited parametrisation
of microbial transmission (e.g. in terms of $\langle T \rangle$) combined
with an insufficient
description of structural heterogeneity in models
(e.g. as in models 1-3) may lead to serious underestimates of the size
of invasions. 
 This highlights the importance of capturing
(i) the essential features of structural heterogeneity that affect microbial
spread and
(ii) the appropriate parametrisation of microbial 
transmission which takes into account the effects of structural 
heterogeneity.
Our results suggest that describing microbial transmission
in terms of the parameters $\lambda_0$ and $\alpha$ (Model 4) is more
appropriate than
using just $\langle T \rangle$. 
Indeed, $\lambda_0$ and $\alpha$ give a better
description of the effects of channel structure on microbial spread and
reveal significant differences in invasion for 
ploughed and unploughed soil
 (microorganisms
with preference for wide channels invade more in ploughed soil, and vice versa~\cite{Supplemental_HeteroSoil}).

To conclude, our work
demonstrates that the shape, size, and interconnection of pores in addition to 
other characteristics influencing the value of the
area exploration parameter $\alpha$ 
(e.g. type of microorganisms)
are the key factors determining
the extend of microbial invasion in soil. 
The interplay between heterogeneity in microbial transmission 
and heterogeneity in the topology of soil networks 
plays a crucial role for biological invasions 
in soil. 

\begin{acknowledgments}
We acknowledge fruitful discussions with V.L. Morales.
CAG gratefully acknowledges support for a BBSRC Professorial Fellowship. LdFC is
grateful to FAPESP (05/00587- 5) 
     and CNPq (301303/06-1 and 573583/2008-0).  MPV
     thanks FAPESP (2010/16310-0) for his post-doc grant.
\end{acknowledgments}

\vspace{-0.6cm}

%%% BibTex bibliography %%%%%%%%%%
%\bibliographystyle{apsrev}
%apsrev4-1.bst     - BibTeX styles for use for Phys. Rev. journals
%apsrev4-1long.bst - Same as above, but shows titles for cited journal articles
%\bibliographystyle{apsrev4-1}
%\bibliography{bibliography}

%

\end{document}